# Anti-pig Antibodies in Swine Veterinarian Serum: Implications for Clinical Xenotransplantation.

Byrne Guerard W[1,2] and McGregor Christopher GA[1,2]

**Affiliations:**

1. University of Minnesota, Twin Cities
   Department of Surgery
   Experimental Surgical Services
   Minneapolis, MN

2. University College London
   Institute of Cardiovascular Sciences
   London WC1E 6BT

Address Correspondence to
Guerard W. Byrne
University of Minnesota, Twin Cities
Department of Surgery
Minneapolis, MN
byrne413@umn.edu

**Financial Support**

This work was supported by Experimental Surgical Services, University of Minnesota and FIOS Therapeutic Inc.

**Conflict of Interest**

Dr. Byrne has no conflicts to declare. Dr. McGregor is founder of FIOS Therapeutics.

**Abbreviations**

| | |
|---|---|
| AASP | American Association of Swine Practitioners |
| AMR | Antibody mediated rejection |
| APC | Allophycocyanin |
| BSA | bovine serum albumin |
| CSFE | carboxyfluorescein diacetate, succinimidyl ester |
| eFluor450 | organic cell proliferation dye eFluor450 |
| Gal | the galactose a 1,3 galactose glycan encoded by the GGTA-1 gene |
| HEK | human embryonic kidney cells |
| HLA | human leukocyte antigen |
| Neu5Gc | N-glycolylneuraminic acid modified glycans dependent on a functional CMAH gene |
| NHP | nonhuman primates |
| PBS | phosphate buffered saline |
| SDa | GalNAc β1,4 (Neu5Ac α2,3) Gal glycan encoded by the B4GALNT2 gene |
| SLA | swine leukocyte antigen |
| TKO | triple knockout pigs not producing the Gal, SDa or Neu5Gc glycans |

**Abstract**

Recent clinical xenotransplantation and human decedent studies demonstrate that clinical hyperacute rejection of genetically engineered porcine organs can be reliably avoided but that antibody mediated rejection continues to limit graft survival. We previously identified porcine glycans and proteins which are immunogenic after cardiac xenotransplantation in nonhuman primates, but the clinical immune response to antigens present in glycan depleted triple knockout (TKO) donor pigs is poorly understood. In this study we use fluorescence barcoded HEK cells and HEK cell lines expressing porcine glycans (Gal and SDa) or proteins (CD9, CD46, CD59, PROCR and ANXA2) to screen antibody reactivity in human serum from 160 swine veterinarians, a serum source with potential occupational immune challenge from porcine tissues and pathogens. High levels of anti-Gal IgM were present in all samples and lower levels of anti-SDa IgM were present in 41% of samples. IgM binding to porcine proteins, primarily CD9 and CD46, previously identified as immunogenic in pig to non-human primate cardiac xenograft recipients, was detected in 28 of the 160 swine veterinarian samples. These results suggest that barcoded HEK cell lines expressing porcine protein antigens can be useful for screening human patient serum. A comprehensive analysis of sera from clinical xenotransplant recipients to define a panel of commonly immunogenic porcine antigens will likely be necessary to establish an array of porcine non-Gal antigens for effective monitoring of patient immune responses and allow earlier therapies to reverse antibody mediated rejection.

**Introduction**

Xenotransplantation is dominated by antibody mediated rejection (AMR) characterized by complement dependent cytotoxicity, antibody-dependent cell cytotoxicity and chronic endothelial cell activation [1, 2]. The bulk of human antibody to porcine tissue is directed to the alpha-Gal antigen (Gal) which is the target of discordant complement dependent hyperacute rejection of pig organs in nonhuman primates (NHPs) [3]. With the development and testing of pigs mutated in the porcine GGTA-1 gene required for Gal synthesis (Gal-KO), it became evident that additional antibodies to non-Gal glycan and protein antigens were present and could contributed to xenograft rejection [4]. The non-Gal glycans include N-glycolylneuraminic acid (Neu5Gc) modified glycans and the SID blood group antigen SDa (GalNAc β1,4 (Neu5Ac α2,3) Gal) [5, 6]. Anti-Neu5Gc antibodies, not present in NHP recipients, have long been recognized in humans as a contributory cause of serum sickness after dosing of patients with animal serum [7]. The role of the SID blood group antigen SDa was identified by screening a porcine endothelial cell expression library with sensitized serum IgG from NHPs after cardiac xenotransplantation [6]. This screen identified the porcine B4GALNT2 gene, the enzyme required of synthesis of the SDa glycan, as the source of a new previously unrecognized immunogenic porcine glycan [8]. SID is a polyagglutinable blood group because most individuals have low levels of anti-SDa IgM which agglutinates red blood cells with high SDa levels [5, 9, 10]. Cells from pigs engineered with mutations to eliminate expression of the Gal, Neu5Gc and SDa glycans, triple knockout pigs (TKO), show greatly reduced human antibody binding, with about 30% of patient samples showing no apparent reactivity [11].

Far less is known about the clinical potential for antibodies induced to porcine protein antigens. Antibodies to human leukocyte antigens (HLA) have been shown to cross reacts with certain subsets of swine leukocyte antigen (SLA) [12, 13]. This suggests that a preformed and potentially induced immune response to swine SLA could contribute to clinical xenograft rejection. Others, using proteomic analysis have reported a broader range of human antibody specificities [14]. Consistent with this observation residual human IgM reactivity to TKO pig cells has been described from a variety of human serum sources [15].

The two recent first clinical cardiac xenotransplants implanting TKO 10-gene modified organs both failed with evidence of apparent AMR, highlighting the critical importance of identifying clinical antibody responses to non-Gal antigens [16]. Early analysis of non-Gal immune responses in NHPs suggested a pan-pig response to porcine proteins [17]. Proteomic studies suggested a range of potential target proteins with some evidence of immunodominant antigens [18, 19]. Using a porcine endothelial cell expression library, we identified a series of pig endothelial cell membrane protein antigens which are immunogenic in NHPs after cardiac xenotransplantation [6]. In the current study, we examined serum samples from swine veterinarians, with potential occupational exposure to swine antigens, to test human antibody reactivity to specific immunogenic porcine glycans and proteins expressed on human embryonic kidney cells (HEK).

**Materials and Methods**

Swine Practitioner Serum

Serum samples collected from swine veterinarians attending the 1999 Annual Meeting of the American Association of Swine Practitioners (AASP) were previously described [20]. For this study an anonymized set of 160 AASP serum samples were tested for reactivity to porcine antigens expressed on human embryonic kidney 293 cells (HEK).

Cell Culture

HEK cells and stably transfected HEK cells expressing the porcine GGTA-1 (HEK-Gal), B4GALNT2 (HEK-SDa), CD9 (HEK-CD9), CD46 (HEK-CD46), CD59 (HEK-CD59), PROC (HEK-PROCR) and ANXA2 (HEK-ANXA2) have been previously described [6, 21, 22]. All HEK cell lines were grown in DMEM media supplemented with 10% fetal calf serum, 1% glutamine, nonessential amino acids, and 110 mg/L sodium pyruvate. Cells for flow cytometry were collected at 80% confluence from T-75 flasks using 0.25% trypsin (Gibco, Fisher Scientific Waltham, MA).

Cell Barcoding and Cell Staining

HEK and HEK cells expressing porcine antigens were labelled with combinations of the vital dyes carboxyfluorescein diacetate, succinimidyl ester (CSFE) (eBioscience CFSE, Fisher Scientific Waltham, MA) and eFluor450 (eBIoscience proliferation dye eFluor450, Fisher Scientific Waltham, MA) as described [23]. HEK cell lines were labelled separately with combinations of CSFE, at final concentrations of 0, 0.01 or 0.15 uM, and eFluor450, at final concentrations of 0, 0.2 and 3.0 uM, producing 9 distinct fluorescence groups (**Figure 1A**). For labelling cells were collected from tissue culture plates, washed and resuspended in 0.5 ml of phosphate buffered saline (PBS) at 4 x $10^6$/ml. Vital dyes diluted in PBS were distributed into 9 separate tubes containing: 1 (0.0 uM CSFE, 0.0 uM eFL450), 2 (0.0 uM CSFE, 0.2 uM eFL450) , 3 (0 uM CSFE, 3.0 uM eFL450) , 4 (0.01 uM CSFE, 0.0 uM eFL450) , 5 (0.01 uM CSFE, 0.2 uM eFL450) , 6 (0.01 uM CSFE, 3.0 uM eFL450) , 7 (0.15 uM CSFE, 0.0 uM eFL450) , 8 (0.15 uM CSFE, 0.2 uM eFL450) , 9 (0.15 uM CSFE, 3.0 uM eFL450). To each tube PBS and 0.5 ml of cells (2 x $10^6$) were added to a final volume of 1ml. Cells were incubated at 37° C for 8 minutes to label and then quenched with PBS containing 1% BSA (PBS/BSA). The cells were pelleted and resuspended in complete DMEM media and incubated at 37° C for 10 minutes prior to washing twice with PBS/BSA. Labelled cells were resuspended in 1 ml of PBS/BSA and equal volumes of each labelled population were combined for staining by human serum.

Flow Cytometry

Flow cytometry measurements were made using a BD Biosciences (San Jose) FACSCanto II flow cytometer equipped with 405 nm, 488 nm and 633 nm lasers. All 9 CSFE and eFluor450 labelled barcoded HEK cell populations (180,000 cells) were mixed and incubated with human serum (diluted 1:10) at 4° C for 45 minutes, washed with PBS/BSA and stained with an APC-conjugated mouse anti-Human IgM (BD Pharmingen). Antibody binding to specific target porcine antigens was measured by gating on each individual barcoded cell population. For each analysis two HEK control populations were present and a secondary only negative control was used. Positive controls included staining with purified human anti-Gal antibody and with serum from highly sensitized NHP cardiac xenotransplantation recipients [18, 24]. The results were analyzed with FlowJo version 10.9.0. A median fluorescence index (MFI) was calculated by dividing the sample Median fluorescence value by the average median fluorescence for all HEK control cells.

Homology Comparisons

All homology comparisons were performed using reference porcine and human protein sequences from the NCBI nucleotide database. Homology and matrix identity was determined using the Uniprot alignment program Clustal Omega 2.1. Supplemental files show the detailed porcine and human homology comparisons for CD46 isoforms (Supplement_1) and for the other protein antigens (Supplement_2).

Statistics

Background staining of HEK control cells was determined as the average median HEK staining of all HEK samples processed that day. The minimal cutoff for positive reactivity to HEK cells expressing porcine glycans/proteins was set to the average HEK stain for that day plus 3 standard deviation. Serum samples with strong anti-protein reactivity had median antibody reactivity in excess of 5SD higher than the HEK control. Quantitative values are presented as means ± standard deviations.

## Results

We adapted a high throughput cell barcoding scheme to detect human antibody binding to control HEK cells and to HEK cells expressing xeno-specific glycans (Gal or SDa) or porcine proteins (CD9, CD46, CD59, PROCR, ANXA2) from 160 AASP veterinarian serum samples (**Figure 1A-D**). Each serum sample simultaneously stained 9 distinct CFSE and eFlouor450 labelled barcoded cell populations consisting of 2 control HEK cell populations and 7 additional HEH cell population each expressing one porcine glycan or protein antigen. It became evident early on that there was no IgG reactivity to porcine proteins so only IgM results are presented.

<u>Glycan Reactivity</u>

All 160 serum samples showed positive IgM binding to HEK-Gal cells which was minimally 3 standard deviations (SD) higher than the average median IgM binding to control HEK cells (Figure 1C and 2A). For HEK-Gal cells the averaged median fluorescence index was 40.5 ± 31.5, indicating an average 40-fold higher level of antibody binding than to the HEK controls. Over half of the samples (89 of 160, 57%) exhibited exclusively anti-Gal IgM binding and no other anti-porcine reactivity. Anti-SDa IgM was detected in 65 serum samples (41%) with an average MFI of 3.9 ± 2.3, approximately 10-fold lower than anti-Gal reactivity (**Figure 1D and 2B**). All of the SDa reactive sera were also positive for Gal with 43 of 65 sera exclusively binding to HEK-Gal and HEK-SDa with no reactivity to any of the protein expressing cells. A total of 132 AASP serum samples (82.5%) showed IgM binding to Gal or Gal and SDa with no evident anti-protein reactivity.

<u>Porcine Protein reactivity</u>

There were 28 samples which exhibited IgM binding to at least 1 protein antigen (**Figure 2C and Figure 3A-C**). Most of these sera (19 of 28), which always showed anti-Gal and or anti-SDa activity, reacted with only 1 protein (**Table 1**). Reactivity to HEK-CD46 was most common (n=24) followed by reactivity to HEK-CD9 (n=8) and HEK-CD59 (n=6). In addition to anti-glycan binding, 7 serum samples bound to 2 proteins, reacting with CD46 and CD9 in 4 samples and CD46 and CD59 in the other 3 samples. One serum sample reacted with three protein antigen (CD46, CD9 and CD59) and one sample bound all 5 HEK cells expressing porcine proteins. Of these 28 serum samples 14 had anti-protein reactivity, for at least 1 protein antigen, which exceeded 5 standard deviations above the control HEK levels (**Figure 2C**).

**Discussion**

Based on successful preclinical studies in NHPs, xenotransplantation has made significant clinical advances with two recent orthotopic clinical heart transplants with promising outcomes performed at the University of Maryland using TKO 10-gene-edited pig hearts [16, 25-27]. There have also been a series of renal and cardiac transplants with various donor genetics into decedent patients [28-30]. The apparent AMR at 35 and 47 days in the TKO 10-gene-edited heart xenograft recipients and similar AMR at 33 days in a single 2 month renal decedent study (Robert Montgomery, personal communications) suggests that antibody responses continue to contribute to clinical xenograft rejection. Although human anti-HLA antibody is known to cross react with some porcine SLA proteins, the polymorphic status of MHC antigens is not unique to xenotransplantation, suggesting that an induced immune response to a range of endothelial cell proteins is likely to develop.

In this report we examine serum samples from 160 swine veterinarians using a novel high throughput barcoded screening assay. This serum was collected from swine veterinarians and represents a population of veterinarians at high risk for occupational exposure to pig antigens and pathogens. Previous analysis of this serum collection showed that these swine veterinarians had an increased risk of occupational exposure and increased potential risk of zoonotic HEV infection [20]. In the current study we found that all samples contained high levels of anti-Gal IgM with about 40% of the samples showing both anti-Gal and anti-SDa IgM reactivity with no binding to pig protein antigens. The anti-SDa IgM was about 10-fold lower than anti-Gal IgM which is consistent with what is known about the SID blood group [5, 9, 10]. In 28 serum samples we detected evidence of IgM binding to one or more of the porcine proteins, predominantly CD46 and CD9 (**Table 1**). The level of antibody binding to porcine proteins was comparable, though much less frequent, to IgM binding to SDa (**Figure 2B and C**). In 14-of-28 serum samples, anti-protein binding was easily detected exceeding 5 standard deviations above average IgM binding to control HEK cells (**Figure 2C**). These results suggest that some degree of occupational immune challenge occurred in a minority of swine veterinarians leading to easily detected IgM reactivity to porcine proteins.

All porcine proteins in this study show some level of amino acid polymorphism compared to their human homologues (**Table 2, Supplement_1 and Supplement_2**). Anti-protein reactivity in the AASP serum samples was primarily directed at porcine CD46 and CD9. CD46 is a type I transmembrane protein with a small carboxy-terminal hydrophobic transmembrane domain and a short cytoplasmic region. The remainder of the protein forms an extensive extracellular domain composed of four short consensus repeats (SCRs). The more distal SCRs 1 - 3 are highly polymorphic with slightly better conservation present in SCR 4. Alternative splicing creates multiple isoforms of CD46 in both pigs and humans, however, there remains significant polymorphism in all isoforms with the amino acid identity ranging from 44 - 48% (**Supplement_1**). The relatively high levels of polymorphism across the extracellular region of CD46 would contribute to the immunogenicity of this protein. CD9 has 4 transmembrane domains and two extracellular loops. The overall amino acid identity between pig and human is high (90.18%) with perfect conservation across the 4 transmembrane domains. The major polymorphisms occur in the second extracellular loop, a region of 85 amino acids with only 75.9% amino acid identity [31]. Similar levels of polymorphism are present in the extracellular domains of GPI-linked CD59 and in the type I membrane protein PROCR [32]. Porcine annexinA2 shows minimal polymorphism and was poorly detected in this study or in previous studies of serum from NHPs [6].

There are several limitations in this study which limit our conclusions. First we have no insight into the route, nature, timing, frequency or level of occupation exposure that the swine veterinarians may have experienced. As such we cannot comment on the quantitative level of immunogenicity for any of the antigens we have studied. The absence of IgG reactivity to any of the proteins studied may suggest that occupation exposure is far less intense than immunization or xenotransplantation. Each of the HEK cell lines is unique and there will be variations in the level of antigen expression from one line to the other. This variation may have quantitative effects on the sensitivity and level of antibody binding, making comparisons between antigens difficult. Despite these limitations, it is clear that substantial antibody binding to several porcine proteins was present in a notable number of serum samples, suggesting that these, and potentially other porcine proteins, may be clinically immunogenic after xenotransplantation.

**Conclusions**

Human antibody responses to immune challenge from vaccination and allotransplant are affected by a large number of factors which, in addition to immunosuppression, may also impact the development of AMR in clinical xenotransplantation [33, 34]. Here we show that a number of swine veterinarians with potential occupational exposure to pig tissues and pathogens have IgM which binds to specific pig proteins which were previously identified as immunogenic after preclinical cardiac xenotransplantation. While the HEK cell lines used in this study may be useful for screening patient serum, this was not a comprehensive survey of pig protein immunogenicity and it is plausible that proteins identified here and other porcine proteins, including SLA antigens, may be immunogenic in clinical xenotransplantation. There are now a very limited set of clinical samples from cardiac xenotransplantation recipients and decedent studies. Careful analysis of these sera will be valuable to identify a set of commonly immunogenic non-Gal antigens. A panel of such antigens, displayed on HEK cells, as single antigen beads or other display formats, would be a foundation for monitoring patient immune responses. In the case of complement or thrombolytic regulators it may be advantageous in xenotransplantation to substitute the human gene for a porcine gene which is found to be chronically immunogenic. This genetic engineering approach and more effective immunosuppression, rejection monitoring and therapy or even B-cell tolerance may be necessary for long-term clinical xenograft survival.

## References

1. McGregor, C.G.A. and G.W. Byrne, *Porcine to Human Heart Transplantation: Is Clinical Application Now Appropriate?* J Immunol Res, 2017.
2. Mohiuddin, M.M., B. Reichart, G.W. Byrne and C.G. McGregor, *Current status of pig heart xenotransplantation.* Int J Surg, 2015. **23**(Pt B): p. 234-239.
3. Platt, J.L., S.S. Lin and C.G. McGregor, *Acute vascular rejection.* Xenotransplantation, 1998. **5**(3): p. 169-175.
4. Kolber-Simonds, D., L. Lai, S.R. Watt, M. Denaro, S. Arn, M.L. Augenstein, et al., *Production of alpha-1,3-galactosyltransferase null pigs by means of nuclear transfer with fibroblasts bearing loss of heterozygosity mutations.* Proc Natl Acad Sci U S A, 2004. **101**(19): p. 7335-40.
5. Stenfelt, L., A. Hellberg and M.L. Olsson, *SID: a new carbohydrate blood group system based on a well-characterized but still mysterious antigen of great pathophysiologic interest.* Immunohematology, 2023. **39**(1): p. 1-10.
6. Byrne, G.W., P.G. Stalboerger, Z. Du, T.R. Davis and C.G. McGregor, *Identification of new carbohydrate and membrane protein antigens in cardiac xenotransplantation.* Transplantation, 2011. **91**(3): p. 287-92.
7. Dhar, C., A. Sasmal and A. Varki, *From "Serum Sickness" to "Xenosialitis": Past, Present, and Future Significance of the Non-human Sialic Acid Neu5Gc.* Front Immunol, 2019. **10**: p. 807.
8. Byrne, G., S. Ahmad-Villiers, Z. Du and C. McGregor, *B4GALNT2 and xenotransplantation: A newly appreciated xenogeneic antigen.* Xenotransplantation, 2018. **25**(5): p. e12394.
9. Duca, M., N. Malagolini and F. Dall'Olio, *The story of the Sd(a) antigen and of its cognate enzyme B4GALNT2: What is new?* Glycoconj J, 2023. **40**(1): p. 123-133.
10. Bird, G.W. and J. Wingham, *Cad(super Sda) in a British family with eastern connections: a note on the specificity of the Dolichos biflorus lectin.* J Immunogenet, 1976. **3**(5): p. 297-302.
11. Martens, G.R., L.M. Reyes, P. Li, J.R. Butler, J.M. Ladowski, J.L. Estrada, et al., *Humoral Reactivity of Renal Transplant-Waitlisted Patients to Cells From GGTA1/CMAH/B4GalNT2, and SLA Class I Knockout Pigs.* Transplantation, 2017. **101**(4): p. e86-e92.
12. Martens, G.R., J.M. Ladowski, J. Estrada, Z.Y. Wang, L.M. Reyes, J. Easlick, et al., *HLA Class I-sensitized Renal Transplant Patients Have Antibody Binding to SLA Class I Epitopes.* Transplantation, 2019. **103**(8): p. 1620-1629.
13. Ladowski, J.M., L.M. Reyes, G.R. Martens, J.R. Butler, Z.Y. Wang, D.E. Eckhoff, et al., *Swine Leukocyte Antigen Class II Is a Xenoantigen.* Transplantation, 2018. **102**(2): p. 249-254.
14. Burlak, C., Z.Y. Wang, R.K. Chihara, A.J. Lutz, Y. Wang, J.L. Estrada, et al., *Identification of human preformed antibody targets in GTKO pigs.* Xenotransplantation, 2012. **19**(2): p. 92-101.
15. He, S., T. Li, H. Feng, J. Du, D.K.C. Cooper, H. Hara, et al., *Incidence of serum antibodies to xenoantigens on triple-knockout pig cells in different human groups.* Xenotransplantation, 2023: p. e12818.
16. Griffith, B.P., C.E. Goerlich, A.K. Singh, M. Rothblatt, C.L. Lau, A. Shah, et al., *Genetically Modified Porcine-to-Human Cardiac Xenotransplantation.* N Engl J Med, 2022. **387**(1): p. 35-44.
17. Buhler, L., Y. Xu, W. Li, A. Zhu and D.K. Cooper, *An investigation of the specificity of induced anti-pig antibodies in baboons.* Xenotransplantation, 2003. **10**(1): p. 88-93.
18. Byrne, G.W., P.G. Stalboerger, E. Davila, C.J. Heppelmann, M.H. Gazi, H.C. McGregor, et al., *Proteomic identification of non-Gal antibody targets after pig-to-primate cardiac xenotransplantation.* Xenotransplantation, 2008. **15**(4): p. 268-76.
19. Gollackner, B., I. Qawi, S. Daniel, E. Kaczmarek, D.K. Cooper and S.C. Robson, *Potential target molecules on pig kidneys recognized by naive and elicited baboon antibodies.* Xenotransplantation, 2004. **11**(4): p. 380-1.

20. Meng, X.J., B. Wiseman, F. Elvinger, D.K. Guenette, T.E. Toth, R.E. Engle, et al., *Prevalence of antibodies to hepatitis E virus in veterinarians working with swine and in normal blood donors in the United States and other countries.* J Clin Microbiol, 2002. **40**: p. 117-122.
21. Byrne, G.W., Z. Du, P. Stalboerger, H. Kogelberg and C.G. McGregor, *Cloning and expression of porcine beta1,4 N-acetylgalactosaminyl transferase encoding a new xenoreactive antigen.* Xenotransplantation, 2014. **21**(6): p. 543-554.
22. McGregor, C.G., G.W. Byrne, Z. Fan, C.J. Davies and I.A. Polejaeva, *Genetically Engineered Sheep: a New Paradigm for Future Preclinical Testing of Biological Heart Valves.* J Thorac Cardiovasc Surg, 2023. **in press**.
23. Lu, M., B.M. Chan, P.W. Schow, W.S. Chang and C.T. King, *High-throughput screening of hybridoma supernatants using multiplexed fluorescent cell barcoding on live cells.* J Immunol Methods, 2017. **451**: p. 20-27.
24. McGregor, C.G., A. Carpentier, N. Lila, J.S. Logan and G.W. Byrne, *Cardiac xenotransplantation technology provides materials for improved bioprosthetic heart valves.* J Thorac Cardiovasc Surg, 2011. **141**(1): p. 269-275.
25. Goerlich, C.E., B. Griffith, P. Hanna, S.N. Hong, D. Ayares, A.K. Singh, et al., *The growth of xenotransplanted hearts can be reduced with growth hormone receptor knockout pig donors.* J Thorac Cardiovasc Surg, 2023. **165**(2): p. e69-e81.
26. Langin, M., T. Mayr, B. Reichart, S. Michel, S. Buchholz, S. Guethoff, et al., *Consistent success in life-supporting porcine cardiac xenotransplantation.* Nature, 2018. **564**(7736): p. 430-433.
27. Mohiuddin, M.M., A.K. Singh, L. Scobie, C.E. Goerlich, A. Grazioli, K. Saharia, et al., *Graft dysfunction in compassionate use of genetically engineered pig-to-human cardiac xenotransplantation: a case report.* Lancet, 2023. **402**(10399): p. 397-410.
28. Porrett, P.M., B.J. Orandi, V. Kumar, J. Houp, D. Anderson, A. Cozette Killian, et al., *First clinical-grade porcine kidney xenotransplant using a human decedent model.* Am J Transplant, 2022. **22**(4): p. 1037-1053.
29. Loupy, A., V. Goutaudier, A. Giarraputo, F. Mezine, E. Morgand, B. Robin, et al., *Immune response after pig-to-human kidney xenotransplantation: a multimodal phenotyping study.* Lancet, 2023. **402**(10408): p. 1158-1169.
30. Moazami, N., J.M. Stern, K. Khalil, J.I. Kim, N. Narula, M. Mangiola, et al., *Pig-to-human heart xenotransplantation in two recently deceased human recipients.* Nat Med, 2023. **29**(8): p. 1989-1997.
31. Yubero, N., A. Jimenez-Marin, M. Yerle, L. Morera, M.J. Barbancho, D. Llanes, et al., *Molecular cloning, expression pattern and chromosomal mapping of pig CD9 antigen.* Cytogenet Genome Res, 2003. **101**(2): p. 143-6.
32. Fukudome, K. and C.T. Esmon, *Molecular cloning and expression of murine and bovine endothelial cell protein C/activated protein C receptor (EPCR). The structural and functional conservation in human, bovine, and murine EPCR.* J Biol Chem, 1995. **270**(10): p. 5571-7.
33. Zimmermann, P. and N. Curtis, *Factors That Influence the Immune Response to Vaccination.* Clin Microbiol Rev, 2019. **32**(2).
34. Heeger, P.S., M.C. Haro and S. Jordan, *Translating B cell immunology to the treatment of antibody-mediated allograft rejection.* Nat Rev Nephrol, 2024.

**Tables**

Table 1. Human IgM Reactivity to Porcine Proteins

| Proteins detected (N) | Serum (N) | CD46 | CD9 | CD59 | PROCR | ANXA2 |
|---|---|---|---|---|---|---|
| 1 | 19 | 15 | 3 | 1 | 0 | 0 |
| 2 | 7 | 7 | 4 | 3 | 0 | 0 |
| 3 | 1 | 1 | 0 | 1 | 1 | 0 |
| 4 | 0 | 0 | 0 | 0 | 0 | 0 |
| 5 | 1 | 1 | 1 | 1 | 1 | 1 |

Table 2. Protein Homology

| Protein | amino acids* | % amino acid identity** |
|---|---|---|
| CD46 | 384 | 46.0 |
| CD9 | 226 | 90.2 |
| CD59 | 123 | 47.5 |
| PROCR | 242 | 70.6 |
| ANXA2 | 339 | 97.9 |

* Values are the number of amino acids in the porcine protein. ** CD46 has multiple isoforms. The values for amino acids and % amino acid identity are representative.

**Figure Legends**

**Figure 1.** Barcoding of the HEK cell series and screening for human anti-pig antibody reactivity. **A.** Cells, 1; HEK control, 2; HEK-Gal, 3; HEK-SDa, 4; HEK-CD9, 5; HEK-CD59, 6; HEK-CD46, 7; HEK-PROCR, 8; HEK-ANXA2, and 9; HEK control are stained with different levels of the vital dyes CSFE (x-axis) and eFluor450 (y-axis) as described in the Materials and Methods. **B.** The barcoded population of HEK cells are combined, stained with veterinarian serum and resolved into 9 distinct cell populations based on fluorescence of the vital dyes. Antibody binding to individual antigens is evident by gating on each individual cell population **C.** All serum samples had anti-Gal IgM with 89 samples (56%) showing only anti-Gal IgM reactivity and no reactivity to other porcine antigens. **D.** Anti-SDa and anti-Gal IgM with no binding to any protein antigens was present in 43 samples (27%).

**Figure 2.** **A** summary of AASP swine veterinarian serum IgM binding to glycan (**A and B**) and protein antigens (**C**). **A.** Shows anti-Gal IgM reactivity for 160 AASP swine serum samples. All samples had anti-Gal reactivity. **B.** Shows IgM binding to SDa for 65 AASP serum samples with antibody binding greater than 3 standard deviations above background. **C.** Shows IgM binding to protein antigens for 28 AASP serum samples with antibody binding greater than 3 standard deviations above background. Note the change in y-axis scale between panel **A** and panels **B and C**. Black filled circles in panels **A - C** are IgM binding to control HEK cells. The dotted line in **B and C** represents 5 standard deviations above the average MFI of all HEK control cells.

**Figure 3.** AASP swine veterinarian serum showing anti-porcine protein IgM binding. **A.** Serum showing IgM binding to Gal, SDa and CD46. **B.** Serum showing IgM binding to Gal, SDa and CD9. **C**. Serum showing IgM binding to Gal, SDa, CD9, CD59, CD46, PROCR and ANXA2. All anti-protein reactivity in panels **A-C** exceeded 5 standard deviations above the HEK background, except for the anti-ANXA2 activity in panel **C**

Figure 1.

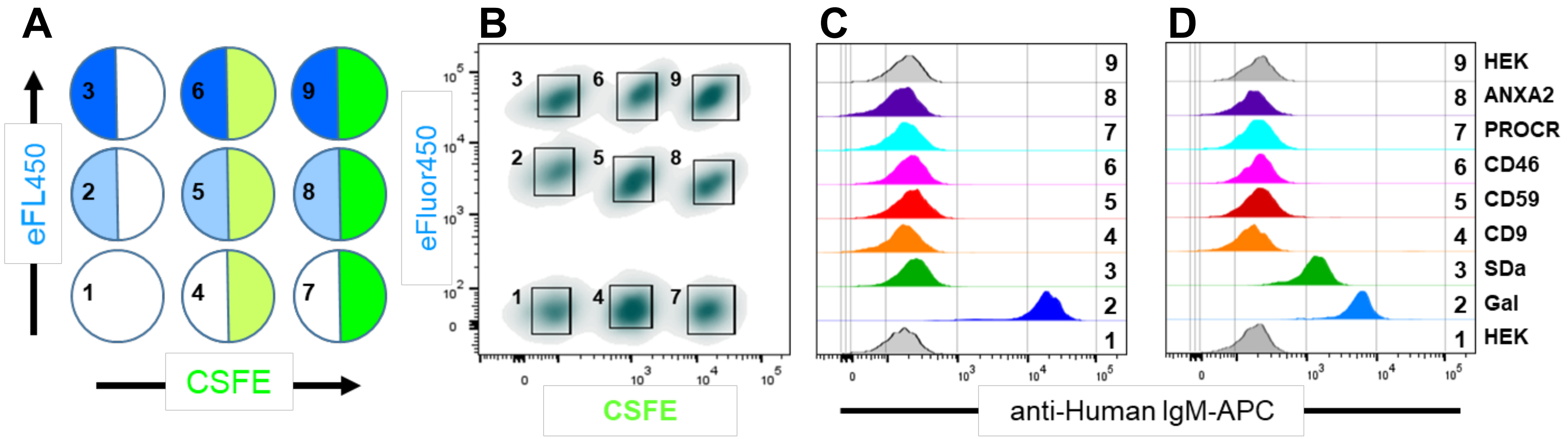

Figure 2.

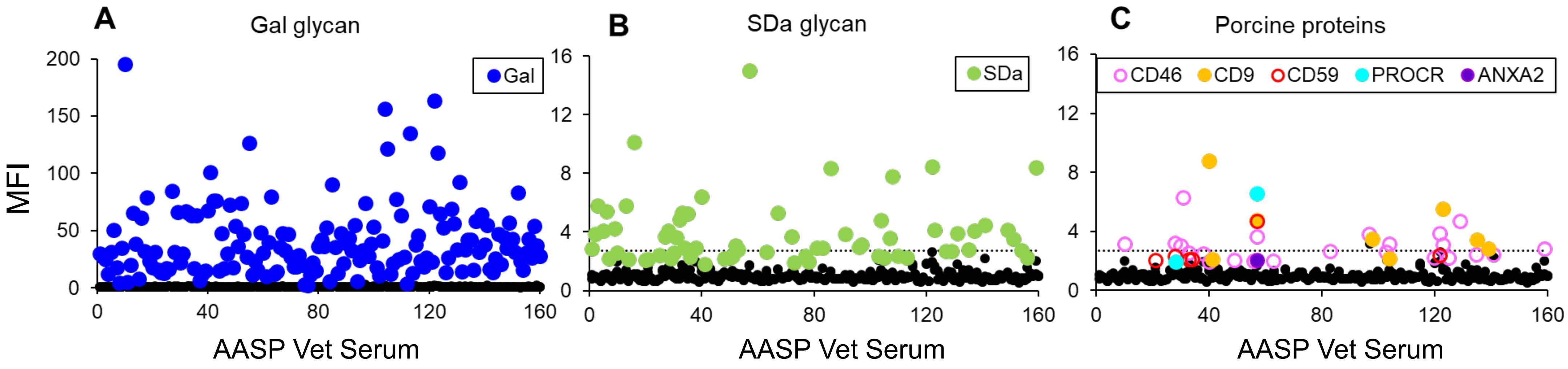

Figure 3.

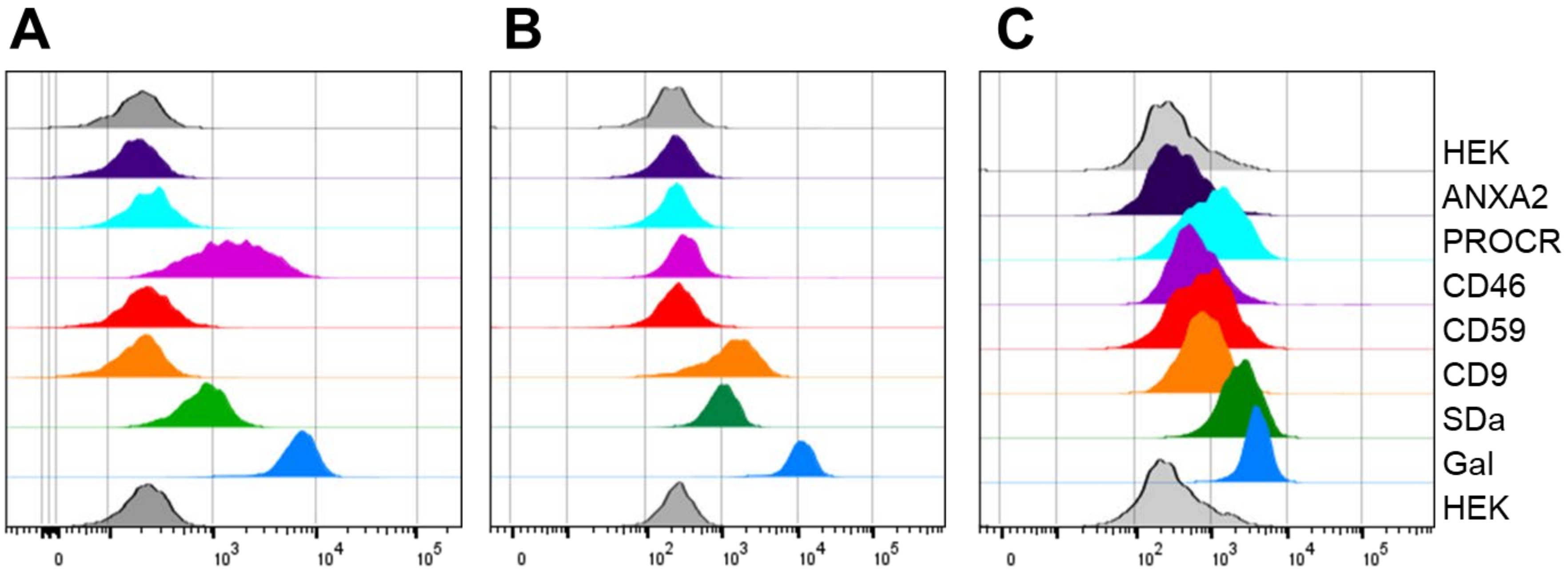

Byrne_Supplement_1

Percent Identity Matrix - created by Clustal2.1

**#Comparing pig and human isoforms of CD46**

**Amino acid identity ranges from 44.35 - 48.57%**

| | | | | | | | | | | | | | | | | |
|---|---|---|---|---|---|---|---|---|---|---|---|---|---|---|---|---|
| 1: PIGXP_020957856.1 | 100.0 | 98.35 | 95.00 | 99.17 | 45.01 | 46.11 | 44.69 | 44.35 | 45.07 | 45.40 | 45.56 | 44.92 | 44.92 | 44.66 | 45.38 | 46.04 |
| 2: PIGNP_999053.1 | 98.35 | 100.0 | 98.62 | 99.17 | 47.29 | 48.57 | 47.04 | 46.65 | 47.46 | 47.87 | 47.32 | 46.63 | 46.63 | 46.31 | 47.09 | 47.87 |
| 3: PIGXP_020957857.1 | 95.00 | 98.62 | 100.0 | 99.45 | 45.69 | 46.83 | 47.47 | 47.09 | 47.88 | 48.27 | 45.43 | 44.79 | 44.79 | 44.54 | 45.25 | 45.91 |
| 4: PIGXP_020957858.1 | 99.17 | 99.17 | 99.45 | 100.0 | 47.29 | 48.57 | 47.04 | 46.65 | 47.46 | 47.87 | 47.32 | 46.63 | 46.63 | 46.31 | 47.09 | 47.87 |
| 5: HUNP_758868.1 | 45.01 | 47.29 | 45.69 | 47.29 | 100.0 | 100.0 | 92.05 | 90.73 | 91.01 | 92.38 | 93.68 | 92.40 | 92.40 | 93.66 | 93.94 | 93.41 |
| 6: HUNP_758860.1 | 46.11 | 48.57 | 46.83 | 48.57 | 100.0 | 100.0 | 94.72 | 92.96 | 92.96 | 94.58 | 95.87 | 94.56 | 94.56 | 95.87 | 95.87 | 95.69 |
| 7: HUNP_758863.1 | 44.69 | 47.04 | 47.47 | 47.04 | 92.05 | 94.72 | 100.0 | 97.29 | 97.81 | 100.0 | 95.80 | 93.31 | 93.31 | 94.75 | 95.26 | 95.69 |
| 8: HUXP_054192621.1 | 44.35 | 46.65 | 47.09 | 46.65 | 90.73 | 92.96 | 97.29 | 100.0 | 99.73 | 100.0 | 94.12 | 94.49 | 94.49 | 94.44 | 93.94 | 95.69 |
| 9: HUNP_758866.1 | 45.07 | 47.46 | 47.88 | 47.46 | 91.01 | 92.96 | 97.81 | 99.73 | 100.0 | 100.0 | 94.12 | 95.79 | 95.79 | 93.94 | 94.21 | 95.69 |
| 10: HUNP_758871.1 | 45.40 | 47.87 | 48.27 | 47.87 | 92.38 | 94.58 | 100.0 | 100.0 | 100.0 | 100.0 | 95.69 | 95.69 | 95.69 | 95.69 | 95.69 | 95.69 |
| 11: HUXP_054192624.1 | 45.56 | 47.32 | 45.43 | 47.32 | 93.68 | 95.87 | 95.80 | 94.12 | 94.12 | 95.69 | 100.0 | 98.87 | 98.87 | 100.0 | 100.0 | 100.0 |
| 12: HUXP_047276857.1 | 44.92 | 46.63 | 44.79 | 46.63 | 92.40 | 94.56 | 93.31 | 94.49 | 95.79 | 95.69 | 98.87 | 100.0 | 100.0 | 97.52 | 98.88 | 100.0 |
| 13: HUMNP_758865.1 | 44.92 | 46.63 | 44.79 | 46.63 | 92.40 | 94.56 | 93.31 | 94.49 | 95.79 | 95.69 | 98.87 | 100.0 | 100.0 | 97.52 | 98.88 | 100.0 |
| 14: HUXP_054192620.1 | 44.66 | 46.31 | 44.54 | 46.31 | 93.66 | 95.87 | 94.75 | 94.44 | 93.94 | 95.69 | 100.0 | 97.52 | 97.52 | 100.00 | 99.73 | 100.0 |
| 15: HUXP_047276850.1 | 45.38 | 47.09 | 45.25 | 47.09 | 93.94 | 95.87 | 95.26 | 93.94 | 94.21 | 95.69 | 100.0 | 98.88 | 98.88 | 99.73 | 100.00 | 100.0 |
| 16: HUNP_758867.1 | 46.04 | 47.87 | 45.91 | 47.87 | 93.41 | 95.69 | 95.69 | 95.69 | 95.69 | 95.69 | 100.0 | 100.00 | 100.0 | 100.0 | 100.0 | 100.0 |

CLUSTAL O(1.2.4) multiple sequence alignment

```
PIGXP_020957856.1       MMAFCALRKALPCRPENPFSSRCFVEILWVSLALVFLLPMPSDACDEPPKFESMRPQVL-   59
PIGNP_999053.1          MMAFCALRKALPCRPENPFSSRCFVEILWVSLALVFLLPMPSDACDEPPKFESMRPQFL-   59
PIGXP_020957857.1       MMAFCALRKALPCRPENPFSSRCFVEILWVSLALVFLLPMPSDACDEPPKFESMRPQVL-   59
PIGXP_020957858.1       MMAFCALRKALPCRPENPFSSRCFVEILWVSLALVFLLPMPSDACDEPPKFESMRPQVL-   59
HUNP_758868.1           --------MEPPGRRECPFPSWRFPGLLLA--AMVLLLYSFSDACEEPPTFEAMELIGKP   50
HUNP_758860.1           --------MEPPGRRECPFPSWRFPGLLLA--AMVLLLYSFSDACEEPPTFEAMELIGKP   50
HUNP_758863.1           --------MEPPGRRECPFPSWRFPGLLLA--AMVLLLYSFSDACEEPPTFEAMELIGKP   50
HUXP_054192621.1        --------MEPPGRRECPFPSWRFPGLLLA--AMVLLLYSFSDACEEPPTFEAMELIGKP   50
HUNP_758866.1           --------MEPPGRRECPFPSWRFPGLLLA--AMVLLLYSFSDACEEPPTFEAMELIGKP   50
HUNP_758871.1           --------MEPPGRRECPFPSWRFPGLLLA--AMVLLLYSFSDACEEPPTFEAMELIGKP   50
HUXP_054192624.1        --------MEPPGRRECPFPSWRFPGLLLA--AMVLLLYSFSDACEEPPTFEAMELIGKP   50
HUXP_047276857.1        --------MEPPGRRECPFPSWRFPGLLLA--AMVLLLYSFSDACEEPPTFEAMELIGKP   50
HUMNP_758865.1          --------MEPPGRRECPFPSWRFPGLLLA--AMVLLLYSFSDACEEPPTFEAMELIGKP   50
HUXP_054192620.1        --------MEPPGRRECPFPSWRFPGLLLA--AMVLLLYSFSDACEEPPTFEAMELIGKP   50
HUXP_047276850.1        --------MEPPGRRECPFPSWRFPGLLLA--AMVLLLYSFSDACEEPPTFEAMELIGKP   50
HUNP_758867.1           --------MEPPGRRECPFPSWRFPGLLLA--AMVLLLYSFSDACEEPPTFEAMELIGKP   50
                                   * * * ** *  *  :* .  *:*:**   ****:***.**:*.

PIGXP_020957856.1       NTTYRPGDRVEYECRPGFQPMVPALPTSSVCQDDNTWSPLQ-EACRRKACSNLPDPLNGQ   118
PIGNP_999053.1          NTTYRPGDRVEYECRPGFQPMVPALPTFSVCQDDNTWSPLQ-EACRRKACSNLPDPLNGQ   118
PIGXP_020957857.1       NTTYRPGDRVEYECRPGFQPMVPALPTSSVCQDDNTWSPLQ-EACRRKACSNLPDPLNGQ   118
PIGXP_020957858.1       NTTYRPGDRVEYECRPGFQPMVPALPTSSVCQDDNTWSPLQ-EACRRKACSNLPDPLNGQ   118
HUNP_758868.1           KPYYEIGERVDYKCKKGYFY-IPPLATHTICDRNHTWLPVSDDACYRETCPYIRDPLNGQ   109
HUNP_758860.1           KPYYEIGERVDYKCKKGYFY-IPPLATHTICDRNHTWLPVSDDACYRETCPYIRDPLNGQ   109
HUNP_758863.1           KPYYEIGERVDYKCKKGYFY-IPPLATHTICDRNHTWLPVSDDACYRETCPYIRDPLNGQ   109
HUXP_054192621.1        KPYYEIGERVDYKCKKGYFY-IPPLATHTICDRNHTWLPVSDDACYRETCPYIRDPLNGQ   109
HUNP_758866.1           KPYYEIGERVDYKCKKGYFY-IPPLATHTICDRNHTWLPVSDDACYRETCPYIRDPLNGQ   109
HUNP_758871.1           KPYYEIGERVDYKCKKGYFY-IPPLATHTICDRNHTWLPVSDDACYRETCPYIRDPLNGQ   109
HUXP_054192624.1        KPYYEIGERVDYKCKKGYFY-IPPLATHTICDRNHTWLPVSDDACYRETCPYIRDPLNGQ   109
HUXP_047276857.1        KPYYEIGERVDYKCKKGYFY-IPPLATHTICDRNHTWLPVSDDACYRETCPYIRDPLNGQ   109
HUMNP_758865.1          KPYYEIGERVDYKCKKGYFY-IPPLATHTICDRNHTWLPVSDDACYRETCPYIRDPLNGQ   109
HUXP_054192620.1        KPYYEIGERVDYKCKKGYFY-IPPLATHTICDRNHTWLPVSDDACYRETCPYIRDPLNGQ   109
HUXP_047276850.1        KPYYEIGERVDYKCKKGYFY-IPPLATHTICDRNHTWLPVSDDACYRETCPYIRDPLNGQ   109
HUNP_758867.1           KPYYEIGERVDYKCKKGYFY-IPPLATHTICDRNHTWLPVSDDACYRETCPYIRDPLNGQ   109
                        :  *. *:**:*:*: *:   :* * * ::*: ::** *:. :** *::*  : ******

PIGXP_020957856.1       VSYPNGDTLFGSKAQFTCNTGFYIIGAETVYCQVSGNVMAWSEPSPLCEKILCKPPGEIP   178
PIGNP_999053.1          VSYPNGDMLFGSKAQFTCNTGFYIIGAETVYCQVSGNVMAWSEPSPLCEKILCKPPGEIP   178
PIGXP_020957857.1       VSYPNGDTLFGSKAQFTCNTGFYIIGAETVYCQVSGNVMAWSEPSPLCEKILCKPPGEIP   178
PIGXP_020957858.1       VSYPNGDTLFGSKAQFTCNTGFYIIGAETVYCQVSGNVMAWSEPSPLCEKILCKPPGEIP   178
HUNP_758868.1           AVPANGTYEFGYQMHFICNEGYYLIGEEILYCELKGSVAIWSGKPPICEKVLCTPPPKIK   169
HUNP_758860.1           AVPANGTYEFGYQMHFICNEGYYLIGEEILYCELKGSVAIWSGKPPICEKVLCTPPPKIK   169
HUNP_758863.1           AVPANGTYEFGYQMHFICNEGYYLIGEEILYCELKGSVAIWSGKPPICEKVLCTPPPKIK   169
HUXP_054192621.1        AVPANGTYEFGYQMHFICNEGYYLIGEEILYCELKGSVAIWSGKPPICEKVLCTPPPKIK   169
HUNP_758866.1           AVPANGTYEFGYQMHFICNEGYYLIGEEILYCELKGSVAIWSGKPPICEKVLCTPPPKIK   169
HUNP_758871.1           AVPANGTYEFGYQMHFICNEGYYLIGEEILYCELKGSVAIWSGKPPICEKVLCTPPPKIK   169
HUXP_054192624.1        AVPANGTYEFGYQMHFICNEGYYLIGEEILYCELKGSVAIWSGKPPICEKVLCTPPPKIK   169
HUXP_047276857.1        AVPANGTYEFGYQMHFICNEGYYLIGEEILYCELKGSVAIWSGKPPICEKVLCTPPPKIK   169
HUMNP_758865.1          AVPANGTYEFGYQMHFICNEGYYLIGEEILYCELKGSVAIWSGKPPICEKVLCTPPPKIK   169
HUXP_054192620.1        AVPANGTYEFGYQMHFICNEGYYLIGEEILYCELKGSVAIWSGKPPICEKVLCTPPPKIK   169
HUXP_047276850.1        AVPANGTYEFGYQMHFICNEGYYLIGEEILYCELKGSVAIWSGKPPICEKVLCTPPPKIK   169
HUNP_758867.1           AVPANGTYEFGYQMHFICNEGYYLIGEEILYCELKGSVAIWSGKPPICEKVLCTPPPKIK   169
                        .   **   ** : :* ** *:*:** * :**::.*.*  **   *:***:**.** :*
```

```
PIGXP_020957856.1      NGKYTNSHKDVFEYNEVVTYSCLSSTGPDEFSLVGESSLFCIGKDEWSSDPPECKVVKCP   238
PIGNP_999053.1         NGKYTNSHKDVFEYNEVVTYSCLSSTGPDEFSLVGESSLFCIGKDEWSSDPPECKVVKCP   238
PIGXP_020957857.1      NGKYTNSHKDVFEYNEVVTYSCLSSTGPDEFSLVGESSLFCIGKDEWSSDPPECKVVKCP   238
PIGXP_020957858.1      NGKYTNSHKDVFEYNEVVTYSCLSSTGPDEFSLVGESSLFCIGKDEWSSDPPECKVVKCP   238
HUNP_758868.1          NGKHTFSEVEVFEYLDAVTYSCDPAPGPDPFSLIGESTIYCGDNSVWSRAAPECKVVKCR   229
HUNP_758860.1          NGKHTFSEVEVFEYLDAVTYSCDPAPGPDPFSLIGESTIYCGDNSVWSRAAPECKVVKCR   229
HUNP_758863.1          NGKHTFSEVEVFEYLDAVTYSCDPAPGPDPFSLIGESTIYCGDNSVWSRAAPECKVVKCR   229
HUXP_054192621.1       NGKHTFSEVEVFEYLDAVTYSCDPAPGPDPFSLIGESTIYCGDNSVWSRAAPECKVVKCR   229
HUNP_758866.1          NGKHTFSEVEVFEYLDAVTYSCDPAPGPDPFSLIGESTIYCGDNSVWSRAAPECKVVKCR   229
HUNP_758871.1          NGKHTFSEVEVFEYLDAVTYSCDPAPGPDPFSLIGESTIYCGDNSVWSRAAPECKVVKCR   229
HUXP_054192624.1       NGKHTFSEVEVFEYLDAVTYSCDPAPGPDPFSLIGESTIYCGDNSVWSRAAPECKVVKCR   229
HUXP_047276857.1       NGKHTFSEVEVFEYLDAVTYSCDPAPGPDPFSLIGESTIYCGDNSVWSRAAPECKVVKCR   229
HUMNP_758865.1         NGKHTFSEVEVFEYLDAVTYSCDPAPGPDPFSLIGESTIYCGDNSVWSRAAPECKVVKCR   229
HUXP_054192620.1       NGKHTFSEVEVFEYLDAVTYSCDPAPGPDPFSLIGESTIYCGDNSVWSRAAPECKVVKCR   229
HUXP_047276850.1       NGKHTFSEVEVFEYLDAVTYSCDPAPGPDPFSLIGESTIYCGDNSVWSRAAPECKVVKCR   229
HUNP_758867.1          NGKHTFSEVEVFEYLDAVTYSCDPAPGPDPFSLIGESTIYCGDNSVWSRAAPECKVVKCR   229
                       ***:* *. :**** :.*****  : *** ***:***:::* .:. **   ********
```

```
PIGXP_020957856.1         YPVVPNGEIVSGFGSKFYYKAEVVFKCNAGFTLHGRDTIVCGANSTWEPEMPQCIKDSKP   298
PIGNP_999053.1            YPVVPNGEIVSGFGSKFYYKAEVVFKCNAGFTLHGRDTIVCGANSTWEPEMPQCIKDSKP   298
PIGXP_020957857.1         YPVVPNGEIVSGFGSKFYYKAEVVFKCNAGFTLHGRDTIVCGANSTWEPEMPQCIKDSKP   298
PIGXP_020957858.1         YPVVPNGEIVSGFGSKFYYKAEVVFKCNAGFTLHGRDTIVCGANSTWEPEMPQCIKDSKP   298
HUNP_758868.1             FPVVENGKQISGFGKKFYYKATVMFECDKGFYLDGSDTIVCDSNSTWDPPVPKCLKVS--   287
HUNP_758860.1             FPVVENGKQISGFGKKFYYKATVMFECDKGFYLDGSDTIVCDSNSTWDPPVPKCLK----   285
HUNP_758863.1             FPVVENGKQISGFGKKFYYKATVMFECDKGFYLDGSDTIVCDSNSTWDPPVPKCLKGPRP   289
HUXP_054192621.1          FPVVENGKQISGFGKKFYYKATVMFECDKGFYLDGSDTIVCDSNSTWDPPVPKCLKVLPP   289
HUNP_758866.1             FPVVENGKQISGFGKKFYYKATVMFECDKGFYLDGSDTIVCDSNSTWDPPVPKCLKVS--   287
HUNP_758871.1             FPVVENGKQISGFGKKFYYKATVMFECDKGFYLDGSDTIVCDSNSTWDPPVPKCLK----   285
HUXP_054192624.1          FPVVENGKQISGFGKKFYYKATVMFECDKGFYLDGSDTIVCDSNSTWDPPVPKCLK----   285
HUXP_047276857.1          FPVVENGKQISGFGKKFYYKATVMFECDKGFYLDGSDTIVCDSNSTWDPPVPKCLKVSTS   289
HUMNP_758865.1            FPVVENGKQISGFGKKFYYKATVMFECDKGFYLDGSDTIVCDSNSTWDPPVPKCLKVSTS   289
HUXP_054192620.1          FPVVENGKQISGFGKKFYYKATVMFECDKGFYLDGSDTIVCDSNSTWDPPVPKCLKVLPP   289
HUXP_047276850.1          FPVVENGKQISGFGKKFYYKATVMFECDKGFYLDGSDTIVCDSNSTWDPPVPKCLKVS--   287
HUNP_758867.1             FPVVENGKQISGFGKKFYYKATVMFECDKGFYLDGSDTIVCDSNSTWDPPVPKCLK----   285
                          :*** **: :****.****** *:*:*: ** *.* *****.:****:* :*:*:*

PIGXP_020957856.1         TDPP--AT------------PGPSHPGPP------SPSDASPPKD----AESLDGGIIAA   334
PIGNP_999053.1            TDPP--AT------------PGPSHPGPP------SPSDASPPKD----AESLDGGIIAA   334
PIGXP_020957857.1         TDPP--AT------------PGPSHPGPP------SPSDASPPKD----AESLDGGIIAA   334
PIGXP_020957858.1         TDPP--AT------------PGPSHPGPP------SPSDASPPKD----AESLDGGIIAA   334
HUNP_758868.1             -------------TSSTTKSPASSASGPRPTYKPPVSNYPGYPKPEEGILDSLDVWVIAV   334
HUNP_758860.1             --------------------------GPRPTYKPPVSNYPGYPKPEEGILDSLDVWVIAV   319
HUNP_758863.1             T-----------------------------YKPPVSNYPGYPKPEEGILDSLDVWVIAV   319
HUXP_054192621.1          SSTKPPALSHSVSTSST--------------TKSPASSASGYPKPEEGILDSLDVWVIAV   335
HUNP_758866.1             -------------TSST--------------TKSPASSASGYPKPEEGILDSLDVWVIAV   320
HUNP_758871.1             ----------------------------------------GYPKPEEGILDSLDVWVIAV   305
HUXP_054192624.1          --------------------------GPRPTYKPPVSNYPGYPKPEEGILDSLDVWVIAV   319
HUXP_047276857.1          STTKS-----------------------------PASSASGYPKPEEGILDSLDVWVIAV   320
HUMNP_758865.1            STTKS-----------------------------PASSASGYPKPEEGILDSLDVWVIAV   320
HUXP_054192620.1          SSTKPPALSHSVSTSSTTKSPASSASGPRPTYKPPVSNYPGYPKPEEGILDSLDVWVIAV   349
HUXP_047276850.1          -------------TSSTTKSPASSASGPRPTYKPPVSNYPGYPKPEEGILDSLDVWVIAV   334
HUNP_758867.1             ----------------------------------------GYPKPEEGILDSLDVWVIAV   305
                                                                  . **      :***  :**.

PIGXP_020957856.1         IVVGVLAAIAVIAGGVYFFHHKYNKKSLPVSVR-VLCWEWQCVSPSHSQQP-- 384
PIGNP_999053.1            IVVGVLAAIAVIAGGVYFFHHKYNKKRSK------------------------ 363
PIGXP_020957857.1         IVVGVLAAIAVIAGGVYFFHHKYNKKRKTEINASYSTYQDKAAAPAE------ 381
PIGXP_020957858.1         IVVGVLAAIAVIAGGVYFFHHKYNKKRSK------------------------ 363
HUNP_758868.1             IVIAIDIFKG------------GRRKGKQMVELNMPLTRL--NQPLQQSREAE 373
HUNP_758860.1             IVIAI---------------------GKQMVELNMPLTRL--NQPLQQSREAE 349
HUNP_758863.1             IVIAIVVGVAVICVVPYRYLQRRKKKGKADGGAEYATYQTKSTTPAEQRG--- 369
HUXP_054192621.1          IVIAIVVGVAVICVVPYRYLQRRKKKGKADGGAEYATYQTKSTTPAEQRG--- 385
HUNP_758866.1             IVIAIVVGVAVICVVPYRYLQRRKKKGKADGGAEYATYQTKSTTPAEQRG--- 370
HUNP_758871.1             IVIAIVVGVAVICVVPYRYLQRRKKKGKADGGAEYATYQTKSTTPAEQRG--- 355
HUXP_054192624.1          IVIAIVVGVAVICVVPYRYLQRRKKKGTYLTDETHREVKFTSL---------- 362
HUXP_047276857.1          IVIAIVVGVAVICVVPYRYLQRRKKKGTYLTDETHREVKFTSL---------- 363
HUMNP_758865.1            IVIAIVVGVAVICVVPYRYLQRRKKKGTYLTDETHREVKFTSL---------- 363
HUXP_054192620.1          IVIAIVVGVAVICVVPYRYLQRRKKKGTYLTDETHREVKFTSL---------- 392
HUXP_047276850.1          IVIAIVVGVAVICVVPYRYLQRRKKKGTYLTDETHREVKFTSL---------- 377
HUNP_758867.1             IVIAIVVGVAVICVVPYRYLQRRKKKGTYLTDETHREVKFTSL---------- 348
                          **:.:
```

**Byrne_Supplement_2**

CLUSTAL O(1.2.4) multiple sequence alignment
**Annexin A2: Identity 97.94%**

```
NP_001005726.1      MSTVHEILCKLSLEGDHSTPASAYGSVKAYTNFDAERDALNIETAIKTKGVDEVTIVNIL  60
NP_004030.1         MSTVHEILCKLSLEGDHSTPPSAYGSVKAYTNFDAERDALNIETAIKTKGVDEVTIVNIL  60
                    ******************** ***************************************

NP_001005726.1      TNRSNEQRQDIAFAYQRRTKKELASALKSALSGHLETVILGLLKTPAQYDASELKASMKG  120
NP_004030.1         TNRSNAQRQDIAFAYQRRTKKELASALKSALSGHLETVILGLLKTPAQYDASELKASMKG  120
                    ***** ******************************************************

NP_001005726.1      LGTDEDSLIEIICSRTNQELQEINRVYKEMYKTDLEKDIISDTSGDFRKLMVALAKGRRA  180
NP_004030.1         LGTDEDSLIEIICSRTNQELQEINRVYKEMYKTDLEKDIISDTSGDFRKLMVALAKGRRA  180
                    ************************************************************

NP_001005726.1      EDGSVIDYELIDQDARDLYDAGVKRKGTDVPKWISIMTERSVCHLQKVFERYKSYSPYDM  240
NP_004030.1         EDGSVIDYELIDQDARDLYDAGVKRKGTDVPKWISIMTERSVPHLQKVFDRYKSYSPYDM  240
                    ****************************************** ******:**********

NP_001005726.1      LESIKKEVKGDLENAFLNLVQCIQNKPLYFADRLYDSMKGKGTRDKVLIXIMVSRSEVDM  300
NP_004030.1         LESIRKEVKGDLENAFLNLVQCIQNKPLYFADRLYDSMKGKGTRDKVLIRIMVSRSEVDM  300
                    ****:******************************************** **********

NP_001005726.1      LKIRSEFKRKYGKSLYNYIQQDTKGDYQKALLYLCGGDD 339
NP_004030.1         LKIRSEFKRKYGKSLYYYIQQDTKGDYQKALLYLCGGDD 339
                    **************** **********************
```

CLUSTAL O(1.2.4) multiple sequence alignment
**CD9: Identity 90.18%**

```
NP_999171.1         MPVKGGTKCIKYLLFGFNFIFWLAGIAVLAIGLWLRFDSQTKSIFEQE--NNNSSFYTGV  58
NP_001400174.1      MPVKGGTKCIKYLLFGFNFIFWLAGIAVLAIGLWLRFDSQTKSIFEQETNNNNSSFYTGV  60
                    ************************************************  **********

NP_999171.1         YILIGAGALMMVVGFLGCCGAVQESQCMLGLFFGFLLVIFAIEIAAAIWGYSHKDQVIKE  118
NP_001400174.1      YILIGAGALMMLVGFLGCCGAVQESQCMLGLFFGFLLVIFAIEIAAAIWGYSHKDEVIKE  120
                    ***********:*******************************************:****

NP_999171.1         VQDFYRDTYNKLKGKDDPQRETLKAIHYALDCCGLMGEVEQLLADICPQRDVLSSLPMKP  178
NP_001400174.1      VQEFYKDTYNKLKTKDEPQRETLKAIHYALNCCGLAGGVEQFISDICPKKDVLETF--TS  178
                    **:**:******* **:*************:**** * ***:::****::***.::  .

NP_999171.1         CPEAIKEVFQNKFHIIGAVGIGIAVVMIFGMIFSMILCCAIRRSREMV       226
NP_001400174.1      CPDAIKEVFDNKFHIIGAVGIGIAVVMIFGMIFSMILCCAIRRNREMV       226
                    **:******:*********************************.****
```

CLUSTAL O(1.2.4) multiple sequence alignment
**CD59: Identity 47.54%**

```
NP_999335.1      MGSKGGFILLWLLSILAVLCHLGHSLQCYNCINPAGSCTTAMNCSHNQDACIFVEAVPPK      60
NP_000602.1      MGIQGGSVLFGLLLVLAVFCHSGHSLQCYNCPNPTADCKTAVNCSSDFDACLITKA-GLQ      59
                 ** :** :*: ** :***:** ********* **:..*.**:*** : ***::.:*   :

NP_999335.1      TYYQCWRFDECNFDFISRNLAEKKLKYNCCRKDLCNKSDATISSGKT------ALLVILL      114
NP_000602.1      VYNKCWKFEHCNFNDVTTRLRENELTYYCCKKDLCNFNEQLENGGTSLSEKTVLLLVTPF      119
                 .* :**:*:.***: :: .* *::*.* **:***** .:   ..*.:       ***  :

NP_999335.1      LVATWHFCL    123
NP_000602.1      LAAAWSLHP    128
                 *.*:* :
```

CLUSTAL O(1.2.4) multiple sequence alignment

**PROCR: Identity 70.59%**

```
NP_001156878.1      MLTTLLPLLPLLFLPGRALCSQKVSDGPRDLRMLQVSYFRSPSQVWYQGNATLGGILTHV   60
NP_006395.2         ---MLTTLLPILLLSGWAFCSQDASDGLQRLHMLQISYFRDPYHVWYQGNASLGGHLTHV   57
                        *  ***:*:* * *:***..*** : *:***:****.* :*******:*** ****

NP_001156878.1      LEGPGHNVTIQQLQPLQEPESWELTKNSLEAYLKEFQGLVQVVHQERGVAFPLIVRCLLG   120
NP_006395.2         LEGPDTNTTIIQLQPLQEPESWARTQSGLQSYLLQFHGLVRLVHQERTLAFPLTIRCFLG   117
                    ****. *.** ***********  *:..*::** :*:***::***** :**** :**:**

NP_001156878.1      CELPPEGSRARVFFEVAVNGSSFMSFQPETASWMARPQAASRVVTYTVEQLNKYNRTRYE   180
NP_006395.2         CELPPEGSRAHVFFEVAVNGSSFVSFRPERALWQADTQVTSGVVTFTLQQLNAYNRTRYE   177
                    **********:************:**:** * * *  *.:* ***:*::*** *******

NP_001156878.1      LREFLQDTCVQYVQKHITTHNLKGSQTGRSYTPLVLGILVGCFIIAGVALCIFLYVGGRR   240
NP_006395.2         LREFLEDTCVQYVQKHISAENTKGSQTSRSYTSLVLGVLVGSFIIAGVAVGIFLCTGGRR   237
                    *****:***********::.* *****.**** ****:***.*******: *** .****

NP_001156878.1      RC 242
NP_006395.2         C- 238
```